%% file: OneD-Memory.tex
\documentclass[amsmath,amssymb,aps]{revtex4-2}

\usepackage{tikz}
\usetikzlibrary{arrows,calc,math}
\usepackage{graphicx}
\usepackage{dcolumn}
\usepackage{bm}
\usepackage{hyperref}
\usepackage{standalone}
\usepackage{mwe}
\usepackage{cleveref}
\usepackage{siunitx}
\usepackage{mathtools}
\usepackage{bm}
\usepackage{esvect}
\usepackage{orcidlink}
\usepackage{soul}

\definecolor{NCSUred}{RGB}{204,0,0}
\definecolor{NCSUk10}{RGB}{242,242,242}
\definecolor{NCSUk25}{RGB}{204,204,204}
\definecolor{NCSUk60}{RGB}{102,102,102}
\definecolor{NCSUk90}{RGB}{51,51,51}
\definecolor{ReynoldsRed}{RGB}{153,0,0}
\definecolor{PyromanFlame}{RGB}{209,73,5}
\definecolor{HuntYellow}{RGB}{253,215,38}
\definecolor{GenomicGreen}{RGB}{125,140,31}
\definecolor{InnovationBlue}{RGB}{66,126,147}
\definecolor{BioIndigo}{RGB}{65,86,161}

\begin{document}

\title{The delicate memory structure of origami switches}

\author{Th\'eo Jules\textsuperscript{1,2,3}\,\orcidlink{0000-0002-0245-3754}}
\email{theo.jules.physics@gmail.com}
\author{Austin Reid\textsuperscript{4}\,\orcidlink{0000-0003-1741-2223}}
\author{Karen E. Daniels\textsuperscript{5}\,\orcidlink{0000-0001-6852-3594}}%
\author{Muhittin Mungan\textsuperscript{6}\,\orcidlink{0000-0002-0352-5422}}
\author{Fr\'ed\'eric Lechenault\textsuperscript{2}\,\orcidlink{0000-0001-6432-5240}}

\affiliation{\textsuperscript{1}Universit\'e de Lyon, Ecole Normale Sup\'erieure de Lyon, Universit\'e Claude Bernard, CNRS, Laboratoire de Physique, F-69342 Lyon, France}
\affiliation{\textsuperscript{2}Laboratoire de Physique de l'\'Ecole Normale Sup\'erieure, ENS, PSL Research University, CNRS, Sorbonne University, Universit\'e Paris Diderot, Sorbonne Paris Cit\'e, 75005 Paris, France}
\affiliation{\textsuperscript{3} Raymond and Beverly Sackler School of Physics and Astronomy, Tel Aviv University, Ramat Aviv, Tel Aviv, 69978, Israel}
\affiliation{\textsuperscript{4}Center for Exploration of Energy and Matter, Indiana University}
\affiliation{\textsuperscript{5}Department of Physics, North Carolina State University}
\affiliation{\textsuperscript{6}Institut f\"{u}r angewandte Mathematik, Universit\"{a}t Bonn, Endenicher Allee 60, 53115 Bonn, Germany}

\date{\today}

\begin{abstract}
While memory effects emerge from systems of wildly varying length- and time-scales, the reduction of a complex system with many interacting elements into one simple enough to be understood without also losing the complex behavior continues to be a challenge.
Here, we investigate how bistable cylindrical origamis provide such a reduction via tunably-interactive memory behaviors. 
We base our investigation on folded sheets of Kresling patterns that function as two-state memory units. 
By linking several units, each with a selected activation energy, we construct a one-dimensional material that exhibits return-point memory.
After a comprehensive experimental analysis of the relation between the geometry of the pattern and the mechanical response for a single bit, we study the memory of a bellows composed of 4 bits arranged in series.
Since these bits are decoupled, the system reduces to the Preisach model and we can drive the bellows to any of its 16 allowable states by following a prescribed sequence of compression and extension.
We show how to reasonably discriminate between states by measuring the system's total height and stiffness near equilibrium.
Furthermore, we establish the existence of geometrically-disallowed defective stable configurations which expand the configuration space to 64 states with a more complex transition pattern.
Using empirical considerations of the mechanics, we analyze the hierarchical structure of the corresponding diagram, which includes Garden of Eden states and subgraphs. We highlight two irreversible transformations, shifting and erasure of the defect, leading to memory behaviors reminiscent of those observed with more complex glassy systems.
\end{abstract}

\maketitle

\section*{Introduction}

Memory is the brain function that enables us to both record past events and remember them in the future.
If we extend this definition to broader considerations, it represents the capacity of any system -- electrical, magnetic, or mechanical  -- to store and recall information about past configurations or events.
Memory-like behaviors appear in many forms all around us.
We regularly experience materials that wear or even break when put under repeated stress.
In this instance, the history of the system effectively modifies its mechanical properties, a process named \emph{aging}~\cite{Bouchaud1992}.
Previous studies, recently reviewed by~\citet{keim_review}, have observed that confined crumpled sheets~\cite{Matan2002} and rubbers~\cite{Diani2009} have a memory of the largest loading; that folded polymeric sheets~\cite{Lahini2017, Bruggen2019, Jules2020} and frictional interfaces~\cite{BarSinai2013, Dillavou2020} have a protocol-dependent dynamical response.
These memory effects are a direct consequence of time-dependent, out-of-equilibrium dynamics.

These types of memory only retain a small amount of information for a short time, with the system eventually relaxing back to an equilibrium state. This property is in conflict with practical storage needs, for which a large amount of data should remain for a long duration. A hard disk drive (HDD) is an excellent storage tool for these exact reasons: its binary data is encoded in a ferromagnetic thin films via magnetic hysteresis~\cite{mayergoyz1986prl, bertotti2006book}.
An electromagnetic head produces a magnetic field able to change and read the disk's local magnetic orientation as a binary bit:  up or down. 
Importantly, the amount of storable data scales inversely with the localized size of the magnetic bit and allows for modern high-capacity information storage.
In essence, the HDD uses the availability of numerous modifiable and readable stable global configurations to memorize a large amount of information.
Following from these ideas, we are motivated to consider mechanical memory, since mechanical systems can also present multiple stable configurations.
Because memory is profoundly linked to the states of the system, its control on scalable systems is also a promising way to design metamaterials, materials with physical properties determined by their structures, with reliable reprogrammable properties~\cite{Chen2021}.

Origami seems a likely candidate for designing this kind of metamaterial. While its origins are in the art of folding paper, origami's definition in a scientific context has expanded to represent all the structures obtained by folding thin sheets, and also to include the fastening of panels (discouraged by artistic practitioners).
The degrees of freedom granted by the design of the creases' pattern, the variety and scalability of the resulting physical properties, and the ease of prototyping make origami a powerful concept for metamaterial fabrication.
Some patterns allow for the folding of large structures into more compact space, generating deployable structures~\cite{Guest1994a, Kuribayashi2006, Schenk2014}.
Others exploit the mechanical interactions between folds to generate a device with multiple stable equilibria~\cite{Hanna2014, Waitukaitis2015, Lechenault2015}.

Importantly for their use in developing mechanical memory, previous studies have shown that cylindrical origamis display mechanical bistability for specific folding patterns~\cite{Jianguo2015, Bos2016, Jianguo2017, reid_geometry_2017, zhang2018bistable, Kidambi2020}.
Moreover, recent research showed that such origamis have potential applications with damping-less impact mitigation~\cite{Yasuda2019}, vibration isolation~\cite{Ishida2017}, crawling robots~\cite{Pagano2017, okuya_crawling_2018, Bhovad2019}, robotic bendable arms~\cite{Wu2021}, mechanical metamaterials~\cite{Novelino2020} or even a property we can exploit, binary memory~\cite{yasuda_origami-based_2017, Zhai2018, Masana2020, Meng2021}.
While the search for mechanical memory using cylindrical origami is not novel, the previous investigations limit their model and analysis to a simplified elastic system with truss elements and forego a crucial component for effective memory, the external non-destructive reading of the configuration.

In this work, we explore the memory capacities of bistable cylindrical origamis generated from Kresling creases' pattern~\cite{Kresling2002}, where the state, folded or deployed, of a barrel, a pair of two mirrored unit cells, encodes for one bit of information.
In the first section of this paper, we detail the characteristics of the pattern and remind the reader of previous geometrical models establishing the existence of two flat-faced solutions for the barrel~\cite{reid_geometry_2017}. 
Then, we concentrate on the mechanical study of a single-barrel module produced from a thin mylar sheet, focusing on the case of uniaxial loading for which the rotation of the ends is suppressed.
We observe a clear structure in the relation between the angles of the folding pattern and the threshold forces for the transition between configurations.
We also confirm that the experimental measurement for the deployed origamis' height is consistent with the geometrical prediction.
Next, we investigate the characteristics of a stack of four modules connected in series.
We describe how the Preisach model defines a quasi-static transition diagram of reachable configurations that we verify experimentally.
We use the study on single modules to highlight how the stack's state is readable from its height at equilibrium and its local elastic response.
Finally, we manually prepare states that we expected to be forbidden and investigate these hidden stable defective-configurations allowed by the material's elasticity. 
We explain how these new configurations modify the transition diagram into a more complex hierarchical graph, presenting properties analogous to amorphous materials, and establish a detailed framework for future analysis.

\section{The Kresling Module as a Mechanical Bit}

\subsection{Origami Bellows}

The term cylindrical origami, which we will also name origami bellows, represents a class of cylindrical structures obtained by
folding flat thin sheets while following precise patterns of creases.
For an abstract ideal origami, the creases are straight and have no hinge bending energy, while the faces are rigid without thickness.
Given these constraints, the whole surface of the folded origami can be divided into triangles.
Consequently, the origami bellows are constrained by Connelly's Bellows Theorem \cite{connelly_bellows_1997}: continuous deformations under the rigid faces constrain must conserve its contained volume.
In other words, for a cylindrical origami to be able to function as a bellows that changes volume under actuation, some elements must be driven away from the idealized flat-faced configuration used to derive its equilibrium state~\cite{Bos2016}. 
Because origamis are folded from thin sheets, the bending properties of the sheets enable such deformations. 
Thus, behaviors such as bistability, snap-through, self-deployment, and plastic fatigue emerge from mechanical frustrations in the overall configuration.

In order to reduce the number of adjustable parameters, a previous study by \citeauthor{reid_geometry_2017} only considered regular patterns fully tiled by a single triangular or trapezoidal shape~\cite{reid_geometry_2017}.
Of the four types of achievable patterns, only the Kresling~\cite{Kresling2002} and the Miura-Ori~\cite{Schenk2013} patterns demonstrate two valid flat-faced configurations, an indicator of mechanical bistability.
The Kresling pattern is characterized by a single obtuse triangle (see \cref{fig:bellowspattern}~A) and can naturally be obtained when torsion is added to the compression of a thin cylinder~\cite{Kresling2002}; it is closely related to the Yoshimura buckling pattern~\cite{Yoshimura1955}, a triangular pattern observed when a thin cylinder is compressed.
Starting from the Kresling pattern, the Miura-Ori pattern is obtained by fusing pairs of triangles with adjacent long-sides into obtuse trapezoids.

Even though the two patterns seem tightly linked, the small difference in tessellation heavily affects the mechanical response of the resulting origami.
For instance, a notable disparity is the evolution of the polygonal base, delimited by the horizontal folds, during the deformation.
On the one hand, a triangular tiling yields a unique cross-sectional geometry for all configurations (see \cref{fig:bellowspattern}~C).
On the other hand, the polygonal base is significantly modified during the deformation when a trapezoidal tiling is considered~\cite{reid_geometry_2017}.
Moreover, a quadrilateral tiling presents more opportunities to make inexpensive small deflections of the faces compared to triangular tiling.
As a consequence, the Kresling bellows appear stiffer under external loading, effectively producing a higher energy barrier between equilibrium configurations, a property which is confirmed experimentally~\cite{reid_geometry_2017}.
Owing to its pronounced bistability and relatively unconstrained geometric demands, the following study will exclusively feature Kresling bellows.

\subsection{Kresling Patterned Module}

\begin{figure}[ht]
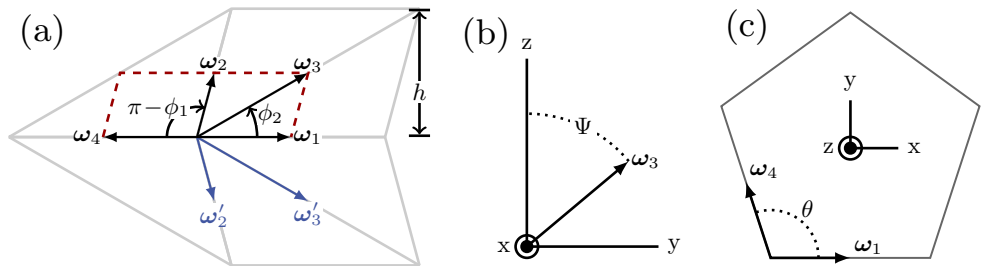

    \centering
    \includestandalone{UnitCellGeometry}
    \caption{Geometry of cylindrical bellows produced with a Kresling pattern. (a) Schematics of the pattern for a single barrel. The unit cell is delimited by the dashed lines,  $\bm{\omega}_1$, and $\bm{\omega}_4$. A Kresling pattern corresponds to $\phi_2 < \phi_1 < \pi/2$. The adjacent unit cell, characterized by $\bm{\omega}_2'$ and $\bm{\omega}_3'$, is a mirror symmetry of the first unit cell along the horizontal fold. (b) Vector $\bm{\omega}_3$ projected in the $YZ$ plane for a deployed configuration. $\Psi$ is the angle between the projection and the z axis. (c) Cross-section in the $XY$ plane delimited by the horizontal folds.}
    \label{fig:bellowspattern}
\end{figure}

The unit cell of the Kresling pattern, shown in \cref{fig:bellowspattern}, is defined by four vectors $\bm{\omega}_i$, with $i=$\numlist{1;2;3;4}, and two angles $\phi_2 < \phi_1 < \pi/2$.
The lengths of both vectors $|\bm{\omega}_1|$ and $|\bm{\omega}_4|$ are identical while both vectors $\bm{\omega}_2$ and $\bm{\omega}_3$ are constrained to reach the same height $\frac{h}{2}$.
Given these restrictions, a unique characteristic length $l_1$, such that $|\bm{\omega}_1| = l_1/2$, is enough to define the pattern's dimensions completely.

The kinematic conditions imposed by the rigid faces hypothesis yield
\begin{equation}
    \sin\Psi = \cfrac{1}{2\tan\cfrac{\pi}{n}\tan\phi_2}\left[1-\frac{\tan\phi_2}{\tan\phi_1}\pm\sqrt{\left(\frac{\tan\phi_2}{\tan\phi_1}-1\right)^2-4\frac{\tan\phi_2}{\tan\phi_1}\tan^2\cfrac{\pi}{n}}\right],
    \label{eq:AngleKresling}
\end{equation}
with $n$ the total number of unit cells within a circumference and $\Psi$ the angle between the $Oz$ axis and the projection of the $\bm{\omega}_3$ vector in the $YZ$ plane (see \cref{fig:bellowspattern}~b). The complete computation is available in~\cite{TheseReid}.

\Cref{eq:AngleKresling} seems intricate at first glance, but the broader mechanical behavior it predicts is straightforward.
The plus or minus before the radical indicates that for any order $n$, there are at most two solutions for this expression and the bellows is at most bistable.
Since the solution only has a physical meaning if $\Psi$ is real-valued, the bistable region's borders is delineated by the constraint $\left|\sin\Psi\right| \leq 1$, which sets a lower boundary $\phi_1 \geq  (\frac{1}{2} + \frac{1}{n})\frac{\pi}{2}$, and a non-crossing constraint imposed as $\phi_1 \leq \frac{\pi}{2}$.
To maximize the bellows' deployability, defined as the difference of the angle $\Psi$ between both solution of the \cref{eq:AngleKresling}, we impose that the collapsed state must be completely flat.
This flat-foldability requires
\begin{equation}
\phi_2 = \phi_1 - \frac{\pi}{n},
\label{eq:Phi1Phi2}
\end{equation}
for all studied bellows.

Finally, we need to account for any rotation induced by the chirality of the unit cell. 
As a way to avoid this issue, we will only consider bellows formed by pairs of mirror-stacked ring that we call a \emph{barrel}.
Since the mirror symmetry yields opposite chirality between the two halves of each barrel, the whole bellows is achiral and will net a zero global internal torsion during folding.

\subsection{Bellows' Production and Experimental Framework}
\label{ssec:ExpFramework}

\begin{figure}[h]
    \centering
    \includegraphics[width=0.7\textwidth]{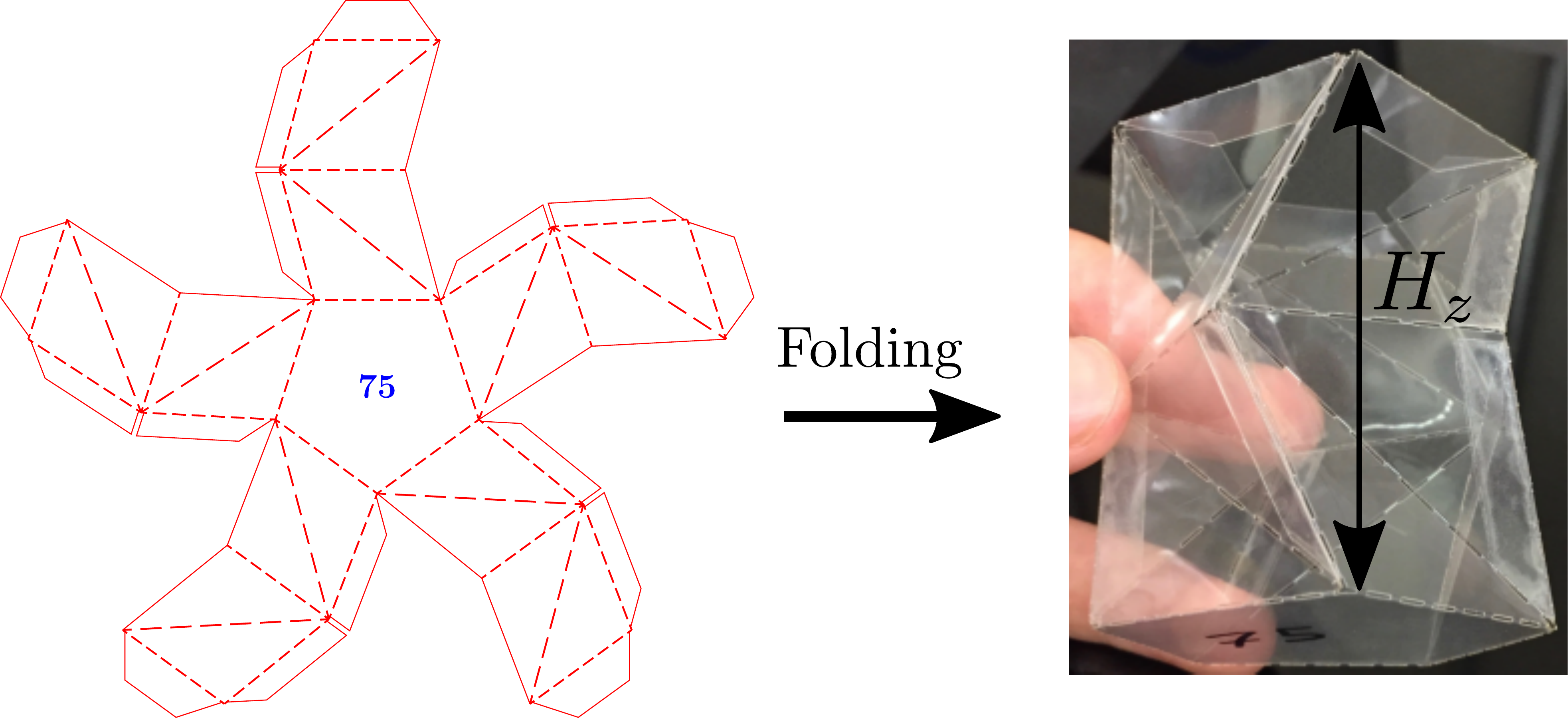}
    \caption{Cutting pattern used to produce a single module with $\ell_1 = 32$ mm, $\phi_1 = 75^\circ$ and $n=5$ next to the resulting bellows after folding. $H_z$ is the bellows' end-to-end height. The drawings are available online~\cite{RepoDrawings}.}
    \label{fig:FoldingBellow}
\end{figure}

The production process for the origami bellows includes two steps.
First, we use a laser cutter to inscribe the desired folding pattern (see~\cref{fig:FoldingBellow}) in \SI{100}{\micro\meter} thick transparent A4 mylar sheets.
We chose $\ell_1 = \SI{32}{\milli\meter}$ to create the largest bellows with a pattern that fits inside an A4 sheet.
Moreover, as a compromise between the difficulty of producing increasingly complex bellows and the size of the configuration space with bistable structures, we chose $n=5$.
We also use the first step to ablate a fraction of each crease's length through a series of perforations, diminishing its local plastic threshold as well as its effective width.
This results in a substantial decrease of the crease stiffness~\cite{Jules2019} without eliminating it entirely and further assures that hinge-bending a fold is easier than bending an adjacent face, resulting in a high energy barrier separating the two stable configurations.
For all the studied samples, we removed \SI{70}{\percent} of each crease's length and the width of each perforation is 0.5 mm.
Then, we used double-sided tape to link neighboring sides of the bellows with predefined tabs.

While our end goal is to study long bellows with multiple barrels and stable configurations, this analysis requires an initial understanding of the properties of a single barrel. 
Hence, we started our investigation by using the star-like pattern presented in~\cref{fig:FoldingBellow}.~(a) to produce \emph{modules}, bellows containing only a single barrel.
Following the geometrical analysis established in the previous section, a module should be bistable as long as the pattern angle $\phi_1$ is between $63^\circ$ and $90^\circ$. In practice, for $\phi_1 < 74^\circ$, the modules have a single stable configuration, while for $\phi_1 > 82^\circ$, the deployed configuration is too rigid and does not properly fold under uniaxial compression. As a result, we limit our study to $\phi_1$ between $74^\circ$ and $82^\circ$.
Please note that the open end is closed with a pentagonal slice of mylar sheet taped to the end tabs.
Then, a simple method to produce longer structure is to stack modules on top of each other and use double-sided tape to connect them.

In order to probe the mechanical properties of the modules and the stacks, we mounted the ends of every sample to plates that can't rotate, themselves attached to a uni-axial testing machine (Instron).
When the ends are allowed to rotate, individual unit cells can actuate \cite{yasuda_origami-based_2017}; by providing rigid clamping, we require that unit cells of opposite chirality collapse in pairs.
The testing device records the end to end height $H_z$ of the bellows and measures the force $F$ applied during deployment-folding experiments.
The exact loading protocol will be given for each experiment.

\subsection{Loading a Single Module}

\begin{figure}[ht]
    \centering
    \includegraphics[width=0.6\textwidth]{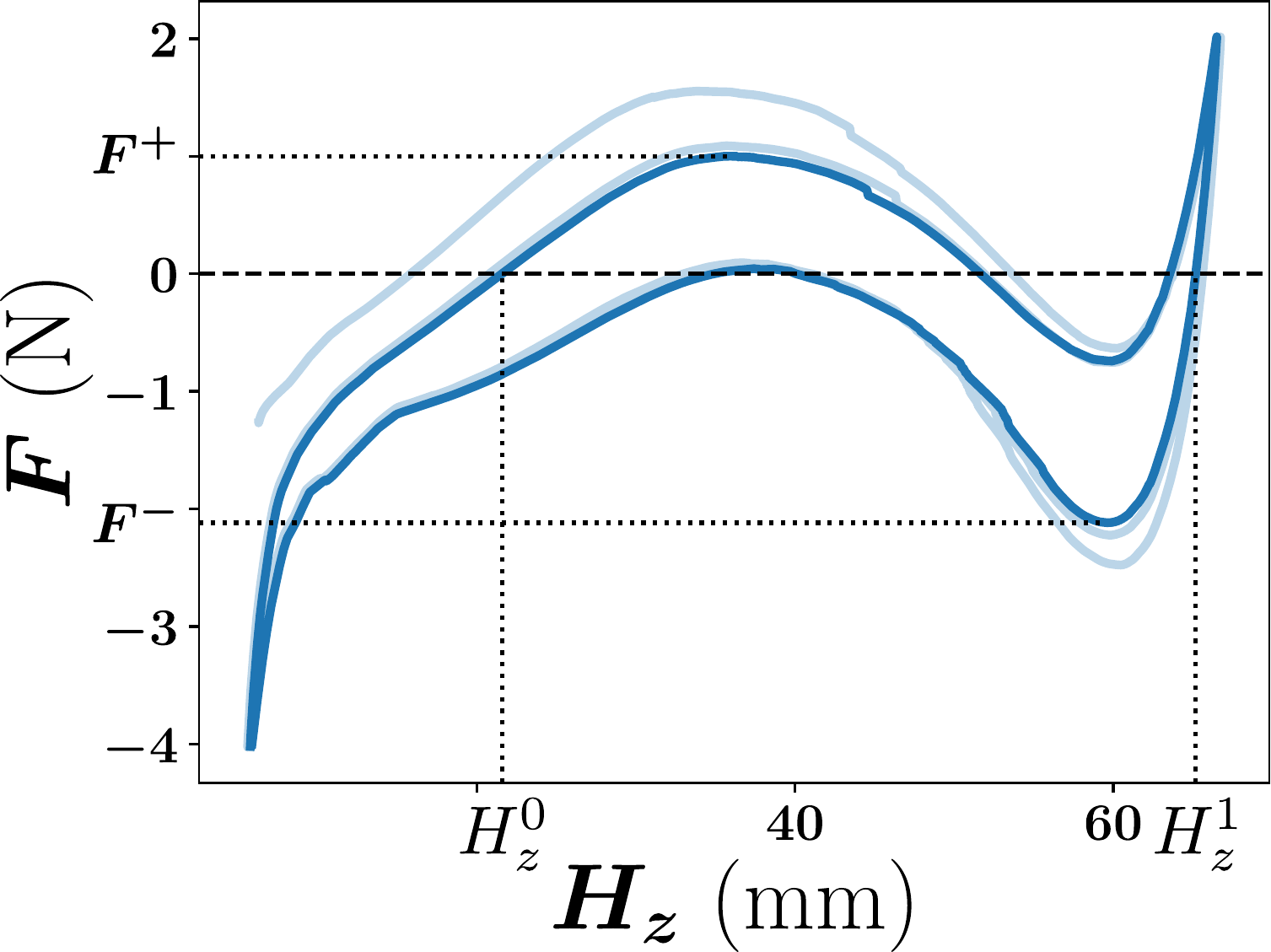}
    \caption{Force-displacement response of a single module with $n=5$, $\phi_1 = 78^\circ$ and $\ell_1$ = 32mm for 3 cycles of deployment-folding. The first two cycles are transparent while the last one is opaque. $F^+ = 1.0$ N and $F^- = -2.11$ N are the thresholds forces between configurations. The heights for folded ($H_z^0$ = 21.6 mm) and deployed ($H_z^1$ = 65.5 mm) configuration are annotated.}
    \label{fig:TypicalResponse}
\end{figure}

Our initial step was to establish the mechanical response of single modules and determine how it depends on the particular folding pattern selected by the choice of $\phi_1$.
To reach that goal we performed cycles of 3 deployment-folding experiments where the module started from, and returned to, the folded configuration.
We ensured that the two limit values for the forces entirely fold and deploy the origami at each cycle while keeping the load low enough not to damage the sample.
A typical force-displacement response is shown in~\cref{fig:TypicalResponse}.
While the response during the first cycle of a newly folded bellows is substantially different, subsequent cycles become roughly repeatable.
We believe that the viscoelastic properties of the creases~\cite{Jules2020} is the main reason behind this behavior. 
Before each experiment, the system has sufficient time to relax to an asymptotic equilibrium; however, this is not possible during an experiment lasting only a few minutes.
As a result, the starting state for the first cycle is unlike any of the other cycles.
This observation is reminiscent of the Kaiser behavior where various physical systems manages to keep a memory of the largest input~\cite{keim_review}.

The cycles present typical bistable behavior with three equilibrium configurations corresponding to the three times the force-displacement curve crosses the $F=0$ line during deployment or folding, where the bistability arises from the flexibility of the faces and the existence of two flat-faces configurations.

Notably, the system displays hysteresis with a different response whether we fold or deploy the bellows, beyond what was reported in the  purely elastic response of a similar truss structure~\cite{yasuda_origami-based_2017, Khazaaleh2021}.
This discrepancy comes from the system's mechanical properties that depend on the history of the sample.
Each crease is characterized by a \emph{rest angle}, an internal characteristic which represent the angle between the faces when the crease is at rest.
When prepared in a folded configuration, every rest angle is required to be small to reach a near-flat folded configuration.
During the deployment, all the creases open and, due to the plasticity of the material, so do the rest angles~\cite{Jules2020}. 
The opposite happens during folding, where the rest angles close.
Consequently, the mechanical equilibrium for the same height $H_z$ will depend on whether the module is being deployed or folded.
An unwanted consequence is that the height of the module in its stable configurations is poorly-defined.
For simplicity, we set the length $H_z^0$ (resp.\ $H_z^1$), the size of the folded (resp.\ deployed) module, to be the minimal (resp.\ maximal) height obtained for $F=0$ during the last cycle. 
We will see later that this choice is in good agreement with the geometrical predictions.

An essential characteristic of a bistable system is the load threshold required to switch from one stable configuration to the other.
The origami bellows' response displays a maximum value for the force during deployment, $F^+$, which is the threshold for the folded to deployed transition. In the same way, we define $F^-$ as the threshold for the deployed to folded transition.
The difference in amplitude between both thresholds comes from the plasticity of the creases and the path of elastic deformation to go from one flat-faced configuration to the other.

\begin{figure}[ht]
    \centering
    \includegraphics[width=\textwidth]{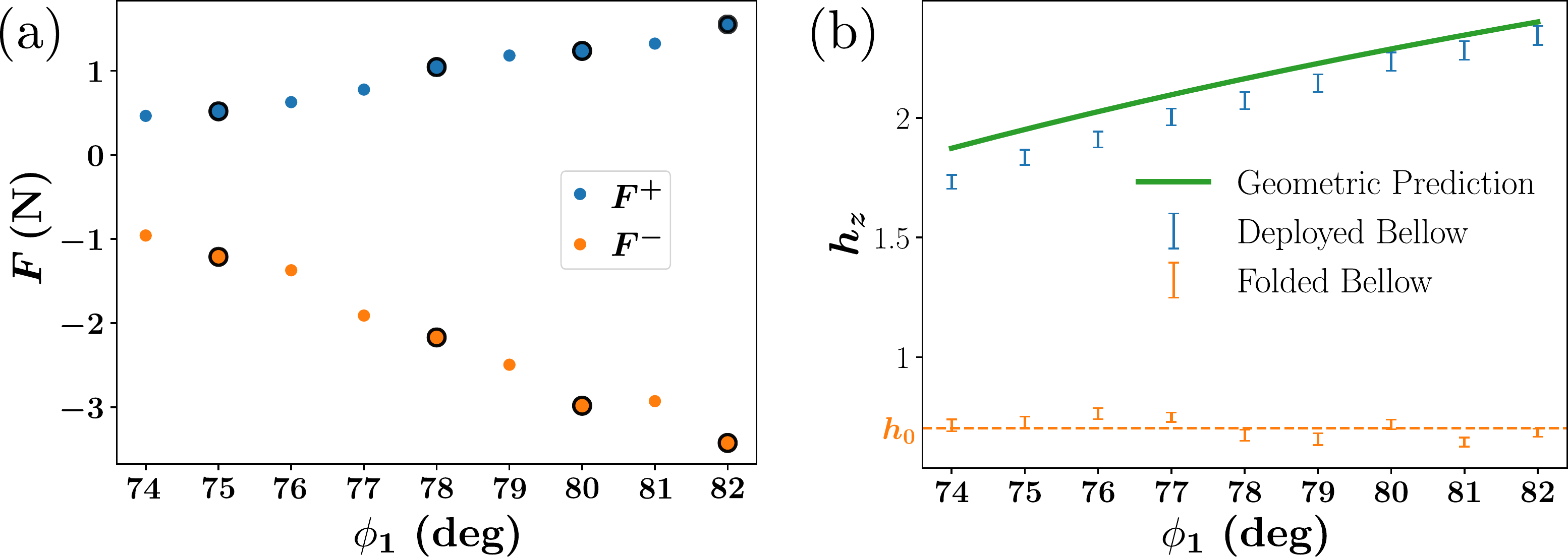}
    \caption{(a) Force thresholds with respect to the pattern angle $\phi_1$. The four highlighted modules are those used in~\cref{subsec:decoupled:bellows}. (b) Normalized height $h_z = \frac{H_z}{\ell_1}$ measured for the folded and deployed configurations. The solid green line corresponds to the geometrical prediction of the deployed height given by \cref{eq:DeployedHeight}. The dashed orange  line corresponds to the mean value $h_0 = 0.70$ measured for the folded height.}
    \label{fig:PatternVariation}
\end{figure}

We tested modules with different angle $\phi_1$ in order to obtain designs generating distinct mechanical response and thresholds.
The measurement of the threshold loads $F^{\pm}$ and normalized heights of the stable configurations $h_z^0 = H_z^0/l_1$ and $h_z^1$ are shown in \cref{fig:PatternVariation}.
During the normalization, we took into account experimental approximations we made while gluing neighbor branches together by taking an interval for the length $\ell_1$ from \SIrange{31}{32}{\milli\meter}.
These considerations leads to the error bars in \cref{fig:PatternVariation}~(b).

The patterns of creases we consider are flat-foldable.
As such, the expected height for the folded configuration is 0.
However, this is not mechanically feasible due to the bellows' finite wall thickness.
Furthermore, we noticed that all of the modules studied showed a similar folded height $H_z^0 \approx 0.65\cdot\ell_1 = \SI{21}{\milli\meter}$, substantially larger than the expected height simply emerging from material thickness $4t \approx 0.4$ mm.
This extra height comes from the small remaining rigidity of the creases, making the stable equilibrium configuration not rigorously flat-faced.
Still, this observation is reassuring with regard to the reproducibility of our test since we used significantly different pattern and obtained a consistent value of $h_z^0$. 
For the remainder of the paper, we set $h_0$ as the normalized height we expect each module to have in the folded state.

In contrast, the deployed height changes dramatically with $\phi_1$.
The theoretical deployed height from the geometrical flat-faced model comes from projecting $\bm{\omega}_3$ on the $z$ axis
\begin{equation}
H_z^1(\phi_1) = \frac{2\ell_1\cos\left[\Psi(\phi_1)\right]}{\cot\left[\phi_2(\phi_1)\right] - \cot\phi_1},
\label{eq:DeployedHeight}
\end{equation}
with $\phi_2(\phi_1)$ the solution of \cref{eq:Phi1Phi2} and $\Psi(\phi_1)$ the negative solution of \cref{eq:AngleKresling}.
Experimentally, this height is measured in our system's state of mechanical equilibrium, which is not rigorously in its flat-faced ideal configuration. Even with slight internal frustration, the measured normalized height of the deployed stable configuration $h_z^1$ stays consistent with the theoretical prediction.

The data show a monotonic dependence of  the value of each threshold force $F^{\pm}$ on the angle $\phi_1$, a behavior that can be understood qualitatively via the simple requirement that, as $\phi_1$ increases, the angle $\Psi$ at the deployed equilibrium decreases. 
In other words, individual bellows become more and more straight for larger $\phi_1$, and the straighter a cylindrical origami is, the more the faces need to bend during the deformation. 
This directly leads to the observed increase of the energy barrier and the threshold forces.

In this section, we established the geometrical and mechanical properties of the Kresling single modules with loading experiments.
We observed that its height in the deployed configuration is in agreement with the geometrical model, while the folded height is independent of the pattern.
Another crucial observation is the relative ordering of both load threshold $F^\pm$ with respect to the pattern.

\section{Bellows as a mechanical memory system}

\subsection{Preisach Model}

\begin{figure}[ht]
    \centering
    \includegraphics[width=0.7\textwidth]{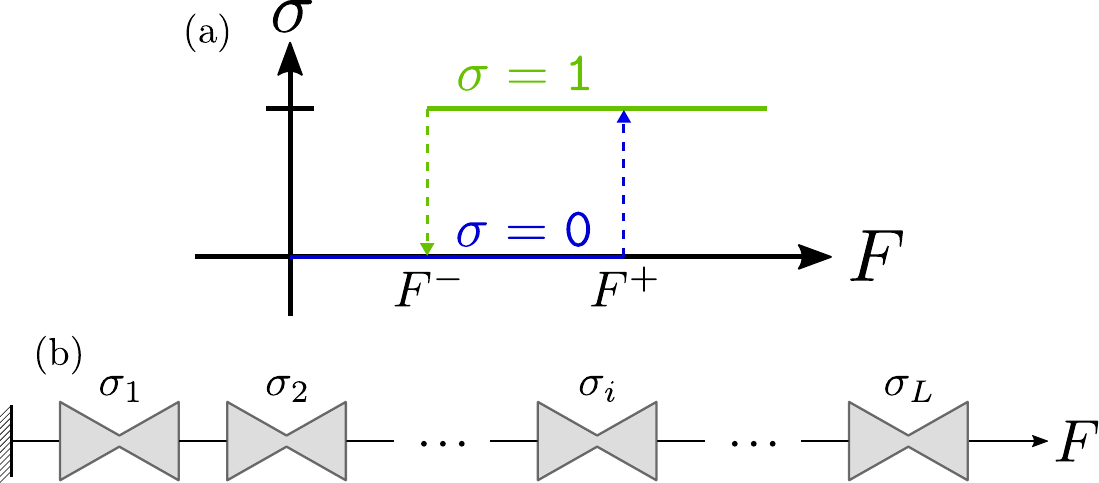}
    \caption{Schematical description of the Preisach model. (a) Transition cycle for an hysteron $\sigma$, element with two stable configurations {\tt 1} and {\tt 0}, with respect to the external loading $F$. $F^+$ (resp. $F^-$) is the loading threshold for the ${\tt 0} \rightarrow {\tt 1}$ (resp. ${\tt 1} \rightarrow {\tt 0}$) transition. (b) Stack of $L$ independent hysterons $\sigma_i$ in series symbolizing a stack of modules.}
    \label{fig:PreisachSchema}
\end{figure}

Using the observed mechanical properties of the modules, we can correctly predict, via Preisach modeling, the quasi-static transition diagram of a stack of modules with different $\Phi_1$, effectively obtaining readable and modifiable mechanical memory.
The Preisach model is a simple mathematical model of systems with hysteresis \cite{Preisach1935, mayergoyz1986prl} which was originally used to describe the hysteretic response in magnetic materials~\cite{bertotti2006book}.
The model consists of a set of $L$ hysteretic elements, called hysterons, each of which can be in one of two states, {\tt 0} or {\tt 1}.
When a hysteron is in state {\tt 0}, it switches to state {\tt 1} as soon as a driving field $F$ becomes larger than a threshold value $F^+$.
To switch back to state {\tt 0}, the driving field must become lower than a second threshold value $F^- < F^+$. 
Thus for $F^- < F < F^+$, the hysteron's state is undetermined and depends on its history, thereby giving rise to hysteresis.
A schematic view of the hysteron transition cycle is presented in \cref{fig:PreisachSchema}~(a).
The theory also assumes that each hysteron couples to the applied external field independently, meaning each transition cycle is unaffected by the other hysterons' states.

Consider an arrangement of $L$ bi-stable modules in series and clamped at one end, labeled $i = 1, 2, \ldots, L$. The hysteretic properties of such systems were studied by  \citet{puglisi2002mechanism}.
When an external force $F$ is applied to the last bellows, an equal force is likewise exerted on each module due to internal equilibrium. 
As depicted in \cref{fig:TypicalResponse}, each bellows $i$ switches from the folded ({\tt 0}) to the deployed ({\tt 1}) configuration and back at threshold forces $F^\pm_i$.
These thresholds' values depend on the pattern via the angle $\phi_1$. 
Suppose the state of each module is independent.
In that case, we can regard this system as a Preisach model, schematized in~\cref{fig:PreisachSchema}~(b), where each bellows is a hysteron.
Therefore, we describe the configuration or microstate of the bellows by an $L$-bit vector, where each bit represents the state of the corresponding bellows.
Finally, we consider very slow changes of the applied driving force so that the response is quasistatic, and we do not need to consider dynamical behaviors.

This model's main hysteresis loop is formed by the transition from the fully folded configuration ${\tt (00\cdots 0)}$ to the completely deployed one ${\tt (11\cdots 1)}$ under increasing force, and back when this force is subsequently reduced.
These two microstates are absorbing: starting from any stable configuration, the fully deployed or folded configuration will be reached for sufficiently large forces. 

The modification of the bellows' microstate due to the external loading can be represented as a transition graph.
The details of such a description have been given in \cite{terzi2020state}.
Here, each vertex corresponds to a configuration that does not transition spontaneously to another configuration and that is compatible with a range of applied forces, the stability interval.
We call such microstate stable.
Depending on the values of the transition thresholds $F^\pm_i$ of the individual modules, not all $2^L$ configurations are stable. 
Transitions between stable configurations occur when the force is increased (resp. decreased) to the upper (resp. lower) limit of a stability interval, and a single hysteron changes its state from ${\tt 0} \to {\tt 1}$ (resp. ${\tt 1} \to {\tt 0}$).
As a result, except for the two absorbing microstates, every stable microstate has two transitions to another stable microstate: a bit either folding or deploying.
Moreover, transitions occur one hysteron at a time, so the Preisach model does not have avalanches.

Among the set of stable states of a Preisach system, one subset is distinct: the set of reachable microstates.
It corresponds to all configurations that can be reached from either of the two absorbing microstates under arbitrary protocols of driving forces.
We call all other, {\em i.e.} non-reachable microstates, ``Garden of Eden'' (GoE) microstates, since these configurations cannot be returned to once the system has landed in a reachable microstate \footnote{The notion of Garden of Eden states goes back to Edward F. Moore \cite{moore1962machine} who credited the coinage to John W. Tukey. It is based on the foundational work of John von Neumann on cellular automata, and in particular how such automata can self-reproduce by creating copies of their states \cite{neumann1966theory}. It turns out that in the analysis of such problems the presence of automata states that one can transit out of but never into takes a central role: these are called Garden of Eden configurations.}.

The transition graph formed by the reachable microstates alone is called the Preisach graph~\cite{terzi2020state}.
It consists of the main hysteresis loop and its sub-loops, which by virtue of the return point memory are organized in a hierarchical structure~\cite{munganterzi2018}.
Moreover, it turns out that the topology of the Preisach graph does not depend on the values of the threshold fields $F^\pm_i$ {\em per se}, but only their relative ordering.
In other words, the crucial property is the sequence in which the individual hysterons change their state when the force is decreased relative to the order when the force is increased~\cite{terzi2020state}.
In particular, if these two orders are equal, there are no GoE microstates and all $2^L$ possible configurations are stable and reachable. 
The Preisach graph in that case for $L = 4$ is depicted in~\cref{fig:DiagrammeDecoupled}~(a), where the folding (resp. deployment) transitions have been marked by black/gray (resp. red/orange) arrows that point to the right (resp. left).

\subsection{Transitions of a stack of individual modules}
\label{subsec:decoupled:bellows}

\begin{figure}[ht]
    \centering
    \includegraphics[width=\textwidth]{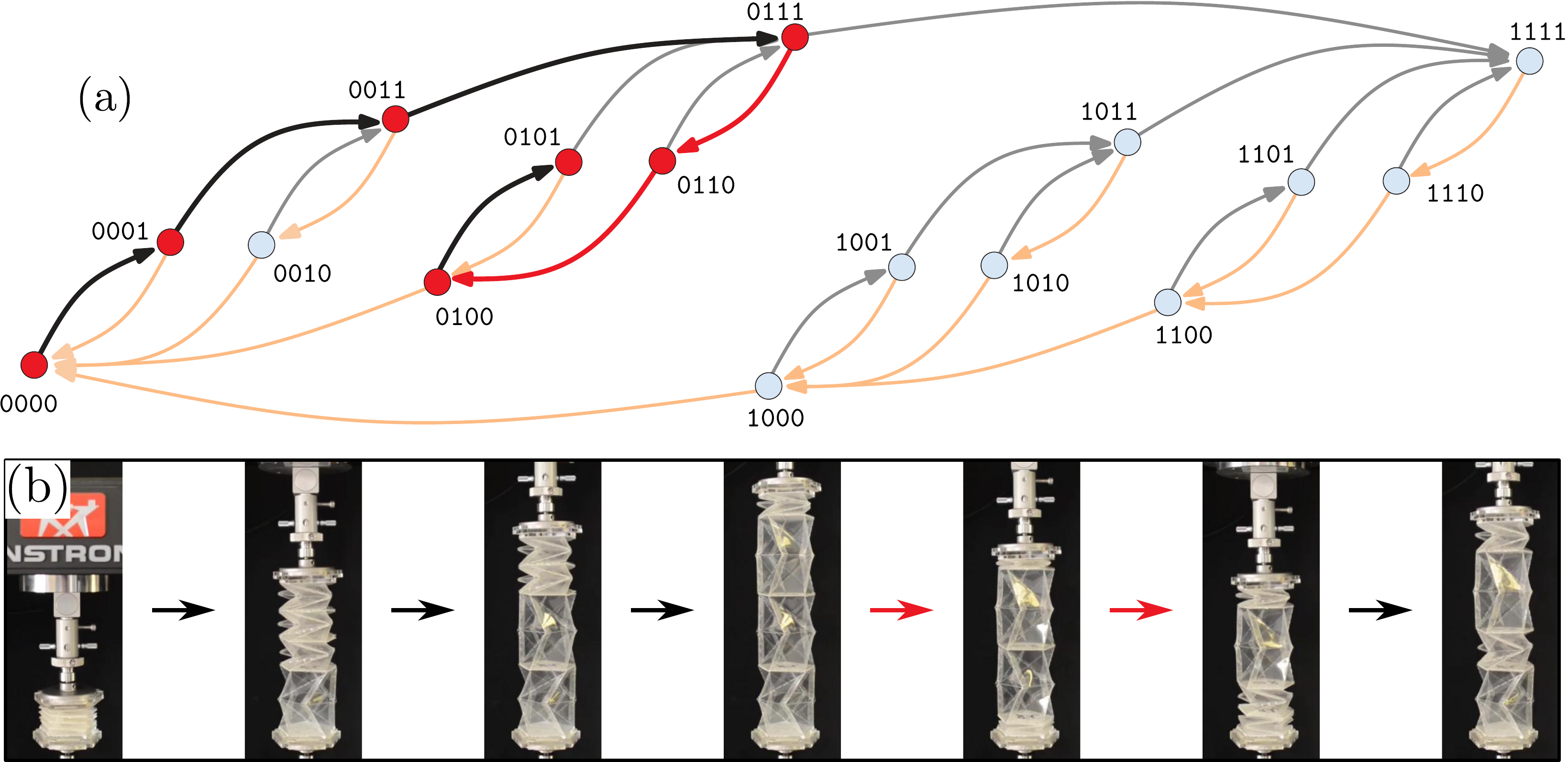}
    \caption{Preisach transition graph for a stack of 4 modules with -- from top to bottom -- $\phi_1 = [82^\circ,80^\circ,78^\circ,75^\circ]$. In the transition graph, each ordered bit {\tt 1} corresponds to a deployed module, and {\tt 0} to a folded one. Each circle is a reachable configuration while each black/gray (or red/orange) arrow corresponds to the deployment (or folding) of one module. Black/red (darker) arrows are a sample pathway to the {\tt 0101} configuration starting from the absorbing microstate {\tt 0000}, as illustrated by the images in (b).}
    \label{fig:DiagrammeDecoupled}
\end{figure}

In order to test the predictions of the Preisach model in an experimental setting, we probed the response of a chain of four linked modules labeled as $i =$~\SIlist{1;2;3;4}{} with respective pattern angles $\phi_1^i =$~\SIlist{82;80;78;75}{\degree}.
From the data collected with single modules in \cref{fig:PatternVariation}~(a), we expect the modules to fold and deploy in the same order: $4 \to 3 \to 2 \to 1$. 
As such, the stack belongs to the case previously described, for which all $2^4 = 16$ configurations are stable and reachable, with the transitions between them specified in the graph presented in \cref{fig:DiagrammeDecoupled}~(a). 
After clamping the stack to our loading device, we explored the transition diagram through quasi-static change of the force and we managed to observe all the transitions predicted by the Preisach model. 
For instance, the ${\tt 0000} \rightarrow {\tt 0101}$ transition is shown in \cref{fig:DiagrammeDecoupled}~(b).
Videos of complete deployment, folding, and the ${\tt 0000} \rightarrow {\tt 0101}$ transition are available in Supplementary Materials.

During our exploration, we examined the origami's response to small deformations around the rest position, corresponding to $F=0$ N, for every configuration.
More precisely, we performed a single loading-unloading cycle between $-0.5$ and $0.5$~N which yields two mechanical characteristics for each microstate, its height and its stiffness.
A typical force-displacement response is shown in the~\cref{app:ResponseStackEq}.

\begin{figure}[ht]
    \centering
    \includegraphics[width=\textwidth]{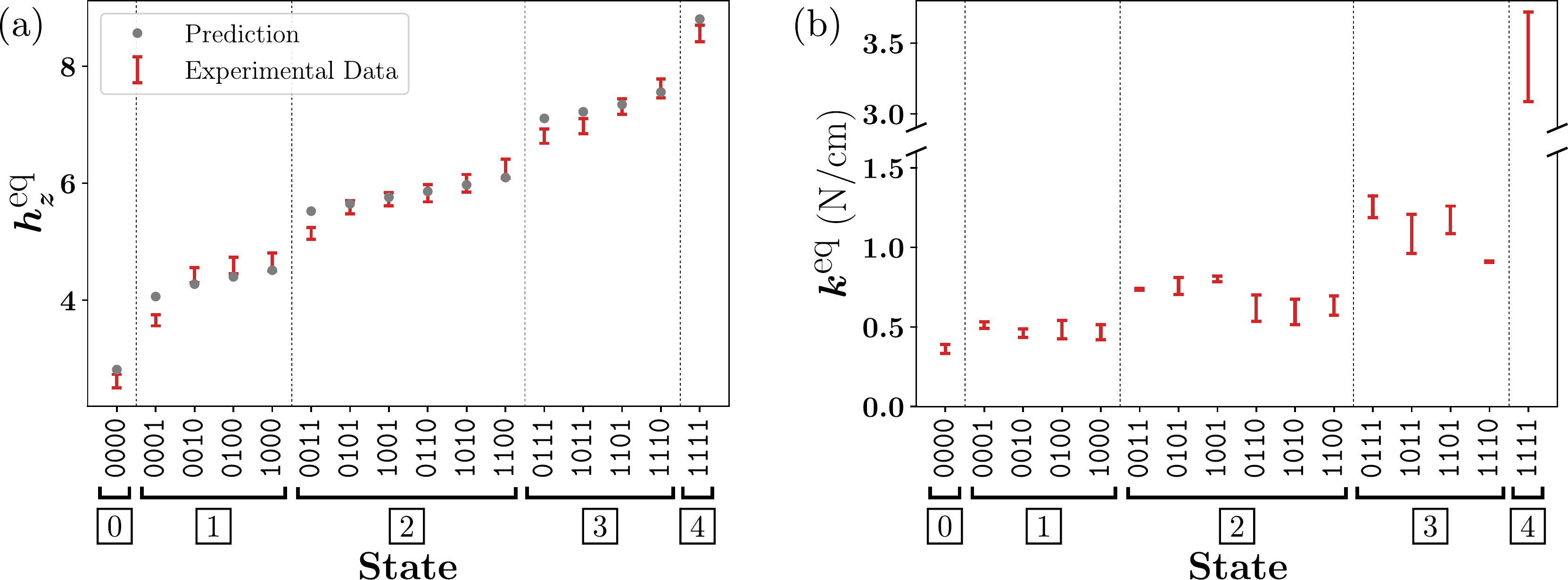}
    \caption{Mechanical properties of the stack of 4 modules. (a) Relative height of each microstate measured experimentally (red error bar) and predicted (gray points) through a method detailed in the text. (b) Measured stiffness $k_{eq}$ of the stack for small deformation in the vicinity of the equilibrium position. The error bars show the difference between loading and unloading. The boxed number \boxed{m} gives the ``magnetization'' macrostate, the number of deployed bellows for each microstate.} 
    \label{fig:StateReading}
\end{figure}

Extending our work on single modules, we compared in~\cref{fig:StateReading}~(a) the recorded relative heights of the configurations at rest $h_z^{\text{eq}}$ to the predicted ones $h_{z\mathrm{,\textrm{ pred}}}^{\textrm{eq}}$. 
The latter is obtained by considering the sum of the height for each module
\begin{align}
h_{z\mathrm{,\textrm{ pred}}}^{\textrm{eq}}(k_1\;k_2\;k_3\;k_4) = \sum_i h_0 + k_i(h_z^1(\phi_1^i)-h_0)
\label{eq:StackHeight}
\end{align}
with $i$ the module number, $\phi_1^i$ the pattern angle and $k_i$ the state of the $i$th module.
For a given module $i$, its deployed height is expected to be the theoretical value $h_z^1(\phi_1)$ given in~\cref{eq:DeployedHeight}, while its folded height is taken as the value $h_z^0$ we measured during single-module loading.
Here again, the theoretical value is in good agreement with the measurements and follows the same ordering.
However, a precise and unequivocal reading of the configuration does not arise from our analysis.
Note that the error bars, coming both from the uncertainty on the value of $\ell_1$ and the interval for $h_z^{\text{eq}}$ found during the mechanical probing near-equilibrium, do not allow for an indisputable identification of the configuration.
Still, if we consider the ``magnetization'' macrostate for the system, corresponding to the total number of deployed modules, we observe a unequivocal step in height for our stack for different macrostates, with smaller differences among its microstates.

A way to break this degeneracy is to examine the configuration's mechanical properties. 
We identify the stiffness of each configuration  using a linear regression of the force-displacement response $F = k^{\text{eq}}(H_z - H_z^{\text{eq}})$ between $-0.1$ and $0.1$~N. 
Once again, we observe a notable difference between loading and unloading, the latter appearing softer. 
We reported our measurements for all microstates in~\cref{fig:StateReading}~(b).
Compared to the height, the data shows that the stiffness is a worse independent identification tool since we observe more overlap between microstates, even for different numbers of deployed bellows.
However, as a complementary tool, it may help discriminate between very similar $h_z^{\text{eq}}$, particularly for larger numbers of units.
For instance, let's look at both configurations {\tt 1001} and {\tt 0110}. 
While they both have comparable lengths, the former appears stiffer. 
So a quick analysis of the mechanical response is enough to differentiate between these two microstates.
Our analysis of a stack of individually-produced modules shows the transition map for the collection of resulting stable configurations is well captured by the Preisach model. 
Moreover, we provide a protocol to read the microstate through their mechanical properties, more precisely their height and apparent stiffness. 

\subsection{Hidden configurations}

\begin{figure}[ht]
    \centering
    \includegraphics[width=\textwidth]{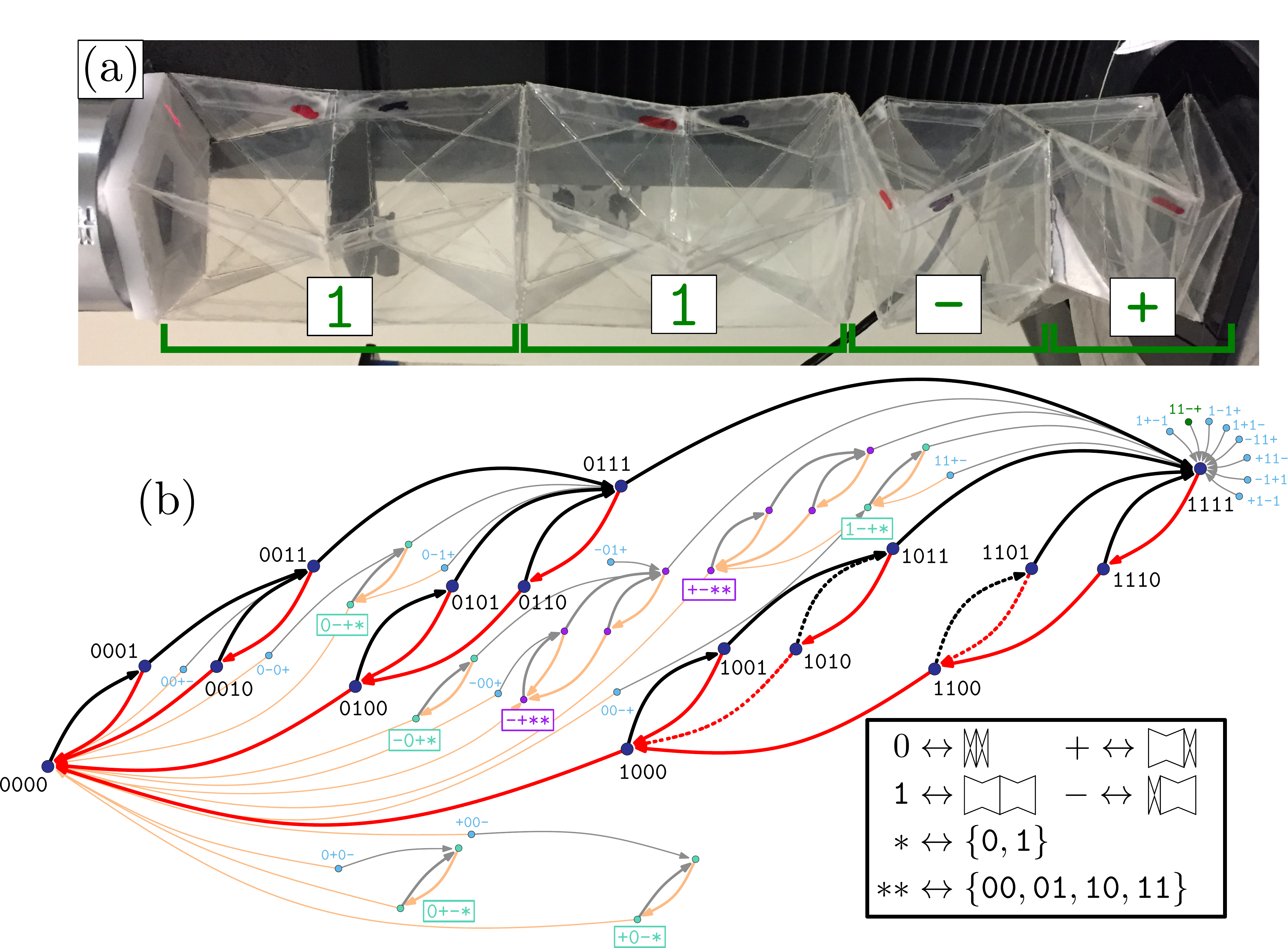}
    \caption{(a) Photo of the defective {\tt 11-+} configuration. (b) Partial directed graph of transitions between stable configurations obtained experimentally. The opaque (or transparent) section is the diagram for reachable (or Garden of Eden, GoE) microstate. The relation between the module's symbol and the state of its halves is schematized in the legend. 2-GoE loops are denoted in cyan and 4-GoE loops in purple. The black and gray (or red and orange) arrows correspond to the deployment (or folding) of a pair of sections observed experimentally. Thin arrows out of the GoE loops represent irreversible ``plastic'' events: \emph{erasure} or \emph{shift}. The dotted arrows represent expected transitions from reachable microstates that we did not observed before the sample's breakdown.}
    \label{fig:DiagramDefects}
\end{figure}

During our investigations, we found that, by carefully twisting parts of a partially compressed bellows, the entire system could be driven into unexpected stable configurations as shown in fig.~\ref{fig:DiagramDefects} (a).
These configurations, obtained by localized manual actuation, present a \emph{defect}: a couple of barrels which are half-folded half-deployed.
More precisely, these barrels have their opposite half folded.
While such defective configurations should be forbidden by the global non-twisting assumption imposed by the boundary conditions, the hidden degrees of freedom due to the material elasticity~\cite{silverberg_origami_2015} allows for opposite halves of slightly different pattern angle to collapse together.

Given these new observations, it is necessary to increase the number of valid microstates we consider for each module.
We will designate half-deployed barrels with new symbols, indicating whether the deployed half faces the stiffer ({\tt +}) or softer ({\tt -}) ends of the bellows.
Since a single defect concerns two barrels, we should consider 54 additional microstates for a 4-barrels bellows, 48 with a single defect, and 6 with two defects (the counting process is detailed in~\cref{app:CountingConf}).
In order to study them, we built a second stack of four modules with pattern angles $\phi_1 = [81^\circ,79^\circ,77^\circ,75^\circ]$, where the three stiffest modules were each made with $\phi_1$ $1^\circ$ smaller compared to the first stack, making them slightly softer.
We only managed to reach configurations with a single defect that maintain a small net twist.
The potential two-defects configurations were not stable, and the system immediately erased one defect.
Moreover, we only managed to observe 39 out of the 48 microstates with a single defect with manual actuation before the sample became too damaged to continue.
All the configurations and their corresponding observed transitions are transcribed in the directed graph in fig.~\ref{fig:DiagramDefects} (b) where we differentiate between the reachable microstates (darker color) and the GoE microstates (lighter color).

Following recent work on the modeling of amorphous solid with directed graphs~\cite{Regev2021}, we will focus our following analysis on the strongly connected components (SCCs) of the transition graph.
The SCCs are the collections of subgraphs where every configuration is reachable from every other configuration.
The largest SCC turns out to be the main graph and -- by definition -- includes only the reachable microstates. 
It is identical to the graph presented in the previous subsection.
Therefore, there is no transition from a reachable microstate to a defective configuration, making them GoE microstates. 
We define $k$-GoE loops as the SCC of the graph formed by $k$ GoE microstates with identical defects.
For instance, 1-GoE loops, which we will also name \emph{isolated} GoE microstate, are single GoE microstate that can not be reached from any other configuration, even defectives ones, through quasi-static transitions.
Our study managed to identify eight isolated GoE microstates, five 2-GoE loops, and two 4-GoE loops.

\subsection{Mechanics and hierarchical structure of the GoE loops}
\label{subsec:mechanics_goe}

The mechanical origin of the GoE loop can be understood as follows.
For any $2^l$-GoE loops, with $l$ a positive integer, we notice the defect does not involve the $l$ softer barrels. 
Therefore, it remains kinetically stuck as long as we keep the external loading small enough.
If this range is enough to actuate the $l$ softer barrels, they will generate a Preisach graph for $l$ independent modules, the $2^l$-GoE loops we observe.
This logic holds for the 4-GoE and 2-GoE loops in the reported graph.
Following this reasoning, for a bellows with $N$ barrels and a single defect, we expect to obtain a total number $n_l$ of $2^l$-GoE loops such that
\begin{align}
n_l = \left (N-l-1 \right ) \, 2^{N-l-1}
\label{eq:nl}
\end{align}
with $l$ a positive integer and $l<N$. 
Here the term $N-l-1$ counts the number of locations at which the other defective barrel can be placed, while the factor $2^{N-l-1}$ accounts for the possibility of the exchange of the two defects and the assignments of states $0$ and $1$ to the $N - l - 2$ inactive barrels.
Since for $l = 1$ and $N = 4$ we have $n_1 = 8$, this implies that there are three additional 2-GoE loops that we did not record in the diagram.

A careful analysis of the graph in~\cref{fig:DiagramDefects}~(c) shows a clear hierarchical structure with nested loops consistent with return-point-memory~\cite{munganterzi2018}.
Any transition from a $2^l$-GoE loop to another loop only occurs when the $l$ softer barrels are either completely deployed or collapse.
The next barrel to be actuated is the $l+1$ softer barrel, which presents a defect.
This leads to two possible outcomes: 
\begin{itemize}
    \item \textbf{\emph{Erasure}}: The defect disappears, and the system shifts to the main cycle.
    This removal is necessary when the total number of deployed or folded modules becomes $N$ and the system transition to the absorbing microstates {\tt 0000} or {\tt 1111}.
    \item \textbf{\emph{Shift}}: The defect moves to a stiffer module.
    This allows additional soft barrels to be freed and contributes to a larger loop.
    In most cases, the defect moves to the next softest barrel. 
    If this barrel were also defective, the system might eliminate the fault and transition to the main cycle.
\end{itemize}
While this analysis successfully predicts most of the transitions observed, some remain inconsistent, such as ${\tt 00-+} \rightarrow {\tt 1-+0}$ for instance.
We argue that these discrepancies come from the aging of the creases during our testing, leading to modifications of the configuration stability.

In our investigation of a stack of modules, we did not observe the last type of transition: the spontaneous \emph{appearance} of a defect.
This transition does not appear because there is insufficient coupling between modules to drive it.
Instead, we observe that the constraints are inadequate to favor the formation of defective configurations over geometrically preferred states.
By introducing a correlation between barrels, we can produce systems for which the formation of defects is beneficial.
An attempt to achieve this is provided in the appendix~\ref{app:CoupledStack}, where we look at a different cutting pattern that provides inter-unit coupling.
In that case, we find that some defective configurations are reachable states.
This observation confirms that tweaking the correlation between barrels is a promising avenue for controlling the transition diagram, but the technicalities are beyond the scope of this study.

\section{Discussion}

In this study, we presented the multiple memory capabilities of cylindrical origami bellows.
We based our analysis on a specific subset of the pattern's configuration space for Kresling cylindrical origamis only using pairs of unit cells, here named barrels.
Our choice guarantees that the folded structures hold three properties: (1) it does not have a global twist under loading, (2) it has two stable configurations, and (3) the pattern's geometry is governed by a single parameter $\phi_1$.
We investigated the mechanical properties of the simplest bellows that we call \emph{modules} and which is composed of only one barrel.
The probing reveals a monotonic relation between the parameter $\phi_1$ and the threshold forces $F_{\pm}$ of transition between the stable microstates.
Moreover, it confirms that a geometrical model based on flat faces correctly predicts the height of the deployed structure.

We exploited all these properties to design a long origami with mechanical memory by stacking modules with different $\phi_1$, and considering the state of each module, deployed or folded, as one bit of memory.
The system requires three distinct features in order to work as effective memory:
(1) Each stable microstate has to remain stable in time.
This is the case for our stack as long as the external loading stays small.
(2) The microstate of the system can be actively changed through external action.
The internal equilibrium assures the force imposed at the extremity of the stack is applied on all the modules.
Furthermore, as we assume them to be independent, and given their mechanical properties, we always obtain a Preisach system where every existing stable configuration is reachable with quasi-static actuation.
This property holds for any number of modules as long as their patterns are all different.
(3) The microstate of the system can be deduced from non-destructive measurements.
From a minute deployment-folding cycle, we showed how the configuration is identifiable with the height and the stiffness of the origami at rest.

More remarkably, the hidden degrees of freedom due to the material elasticity~\cite{silverberg_origami_2015} allow for new stable configurations that a zero global twisting should forbid.
These defective configurations create a more convoluted diagram with Garden of Eden states and irreversible transitions.
Our experimental study shows a hierarchical structure with the graph's strongly connected components induced by the relative mechanical response of the modules.
Notably, the size of any defective loop depends on the stiffness of the softest defective bit of memory which irreversibly increases with the highest load applied to the system during shifting events.
While the model for our system remains simple, we recover behavior typically observed in amorphous materials like the memory of the largest input~\cite{Matan2002, Diani2009} or irreversible plastic events with internal rearrangement due to external stress~\cite{Baret2002}.

From an engineering standpoint, the memory capabilities of the presented cylindrical origamis, in addition to its capacities to tune its mechanical properties, make it an interesting reprogrammable metamaterial in line with recent work on very similar origami structures~\cite{Novelino2020}.
In this paper, we only considered systems composed of four barrels.
In principle, larger systems with distinct reachable states are obtainable since degeneracy only appears if two or more barrels have identical pattern angles $\{\phi_1, \phi_2\}$.
In practice, the material we use and our fabrication method limit the scalability to only a few barrels before a precise reading becomes impossible.
Developing a method to reduce measurement error is essential to make the origami bellows a tractable memory system.
We believe that the observed hysteresis comes from the non-elastic response of the creases, which manifests as changes of their rest angle.
Consequently, we expect new production schemes that reduce creases stiffness to diminish the experimental error and facilitate state reading.
Moreover, long deployed bellows might prefer to buckle out of the central axis during compression instead of folding barrels, a behavior already exploited to create origami octopus-like bendable arms~\cite{Wu2021}.
We also expect longer bellows to generate stable configurations with more defects.
The internal strain in defective configurations is related to the internal twist per unit length.
By increasing the number of barrels, the internal strain generated by any single defect is reduced, which should support the introduction of additional folding defects.

We conclude by pointing out that our works follow a very recent surge of interest for the Preisach model and its generalizations in order to explain the complex transition diagram of multistable systems by introducing interactions between hysterons~\cite{Lindeman2021, Keim2021, MartinvanHecke2021, Shohat2021}.
While our barrel-based approach considers independent hysterons, the existence of defective configuration highlight that each barrel is separable into two mirror halves, which are both bistable.
Consequently, another way to analyze our system is to divide it and consider each half-barrel as a hysteron.
The condition of no-rotation imposed at the ends creates correlations between hysterons and favors transitions where hysterons of opposite chirality transition together.
We leave a thorough investigation of how these correlations precisely relate to the folding pattern and crease properties for future works.

\section*{Acknowledgements}
T.J. thanks Pedro Reis for interesting discussions. 
Research by K.E.D and M.M. was supported in part by the National Science Foundation under Grant No. NSF PHY-1748958. 
M.M. was funded by the Deutsche Forschungsgemeinschaft (DFG, German Research Foundation) under Projektnummer 398962893, the Deutsche Forschungsgemeinschaft (DFG, German Research Foundation) - Projektnummer 211504053 - SFB 1060, and by the Deutsche Forschungsgemeinschaft (DFG, German Research Foundation) under Germany’s Excellence Strategy - GZ 2047/1, Projekt-ID 390685813. 

\bibliography{OneD-Memory}
\newpage

\appendix
\section{Response of the stack near equilibrium}
\label{app:ResponseStackEq}

\begin{figure}[ht]
    \centering
    \includegraphics[width=0.5\textwidth]{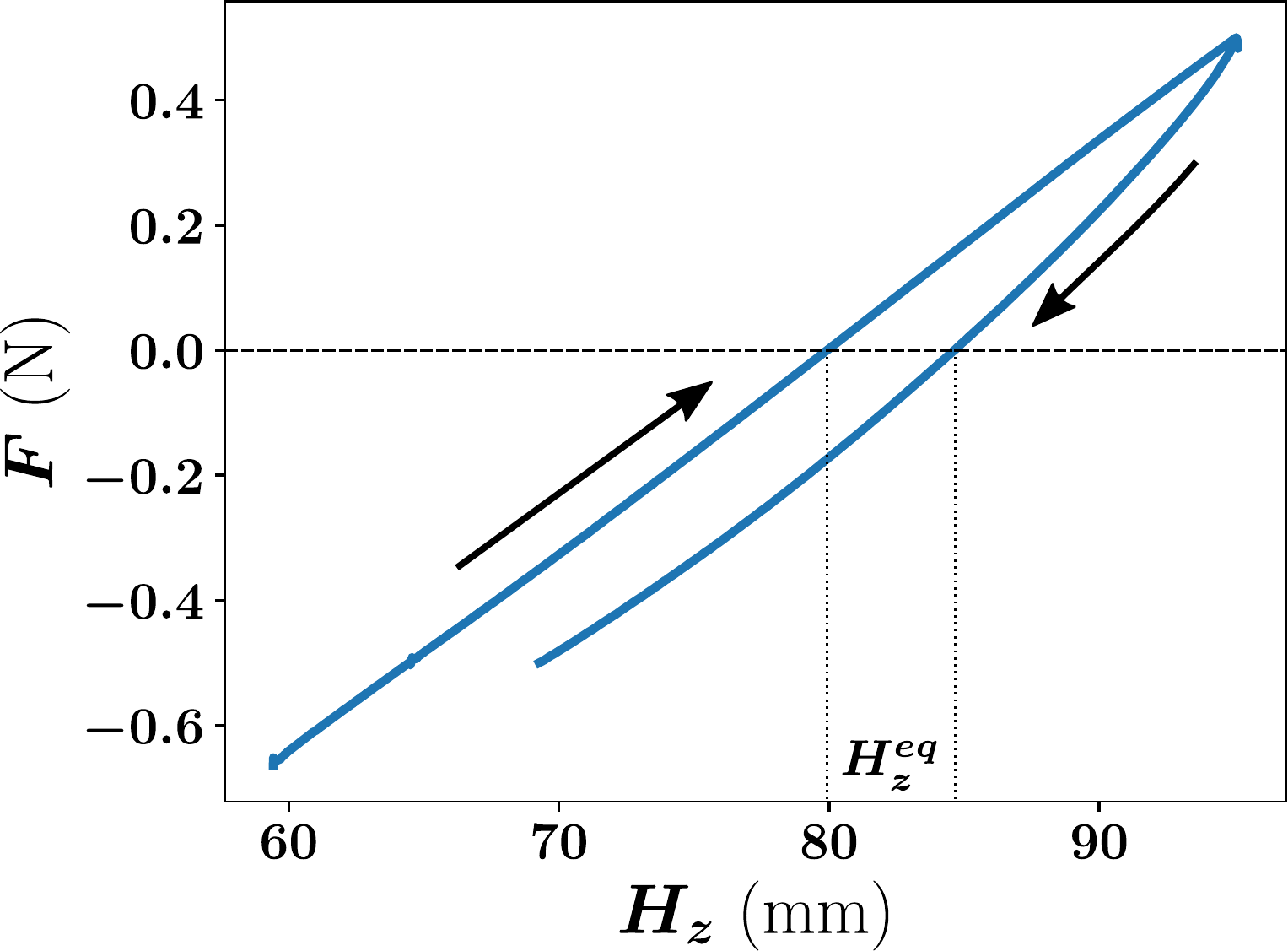}
    \caption{Force-displacement response of the stack of 4 modules studied in~\cref{subsec:decoupled:bellows} near the rest position for the configuration 0000. The dotted lines represent the limit values of $H_z^{eq}$ we used in~\cref{fig:StateReading}}
    \label{fig:SmallResponse}
\end{figure}

We probed the elastic response to small deformation around equilibrium of each reachable state for the stacks of four modules with $\phi_1 = [82^\circ,80^\circ,78^\circ,75^\circ]$ presented in~\cref{subsec:decoupled:bellows}.
To do so, we started from a loading $F \approx -0.5$ N and carried one cycle of extension-compression as shown in~\cref{fig:SmallResponse}. 
We used linear regression on a small region around $F=0$, approximately between $-0.1$ and $0.1$ N, during both extension and compression to set the two limit values for the stiffness $k^{eq}$ used in~\cref{fig:StateReading}. 

\newpage

\section{Number of configuration with \texorpdfstring{$k$}{TEXT} defects for \texorpdfstring{$N$}{TEXT} barrels}
\label{app:CountingConf}

Employing the convention taken in the paper, we consider a defect as a pair of ${\tt +}$ and ${\tt -}$ barrels.
As a result, a defect always concerns two barrels and necessarily $N \geq 2k$, with $N$ the total number of barrels.
Starting from this postulate, counting the number of configurations $n^N_k$ with $k$ defects for a system with $N$ barrels is straightforward.
First, we need to account for all the combinations of barrels concerned by the defects leading to $\binom{N}{2k}$ possibilities.
Then, for all the defectives barrels, half must be in the state ${\tt +}$ while the other half must be in the state ${\tt +}$, leading to $\binom{2k}{2}$ possibilities.
Finally, each non-defective barrel has two state, ${\tt 0}$ or ${\tt 1}$, generating $2^{N-2k}$ possibilities.
As we multiply all these contributions, the total number of configurations with $k$ defects is
\begin{align}
n^N_k = \binom{N}{2k}\binom{2k}{k} 2^{N-2k} = \frac{N!}{k!^2(N-2k)!}2^{N-2k}.
\label{eq:nbr_conf}
\end{align}

The system we studied has $N=4$ barrels and may have up to $k=2$ defects.

The total number of configurations should simply be the sum of configurations realizable with $k$ defects:

\begin{align*}
    n &= \sum\limits_{k=0}^{2} n_k^4\\
    &= 16+48+6\\
    &=70
\end{align*}

However, since we did not observe any configurations with two defects, we consider a total of $n_0^4 + n_1^4 = 64$ configurations.

\section{Case of a long bellows with coupled barrels}
\label{app:CoupledStack}

\begin{figure}[ht]
    \centering
    \includegraphics[width=0.8\textwidth]{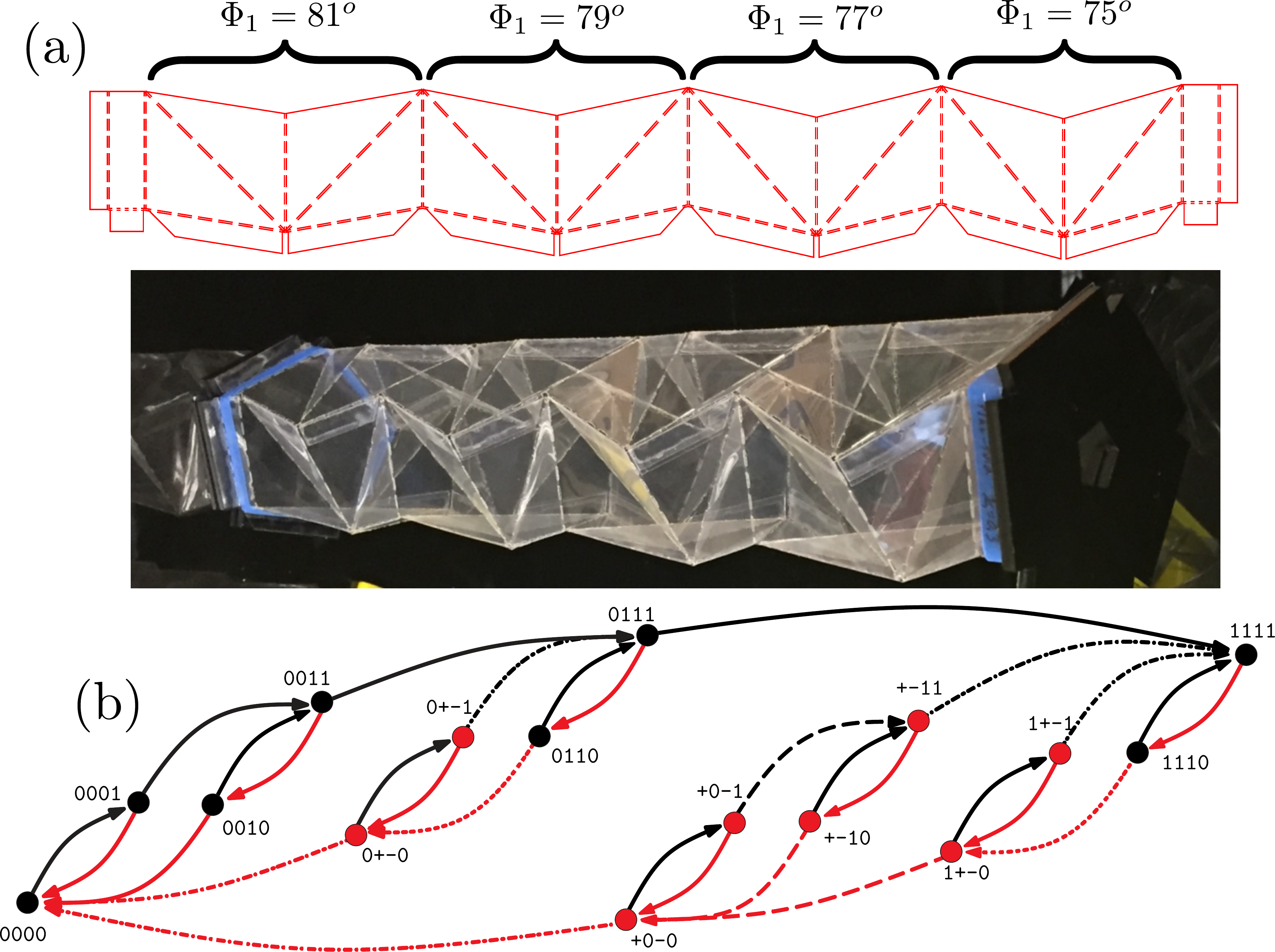}
    \caption{(a) Cutting pattern for a single branch of a 4-barrels stack with $\ell_1 = 32$ mm, $\phi_1 = [81^\circ,79^\circ,77^\circ,75^\circ]$ and the resulting stack for $n=5$ after folding. The drawings are available online~\cite{RepoDrawings}. Below is a photo of the produces bellows for $n=5$ in the {\tt 1111} configurations.
    (b) Observed transition diagram. The black (red) circles represent normal (defective) configuration. A dotted arrow correspond to the \emph{appearance} of a defect, a dashed arrow its \emph{shift} and a doted-dashed arrow its \emph{erasure}.}
    \label{fig:DiagramCoupled}
\end{figure}

While developing our production schemes, we also looked alternative solutions for generating multi-barrel bellows:  a single long cutting pattern rather than preparing individual modules and connecting them afterward.
We extended the design of the branches for the star-like pattern to include multiple barrels.
However, we cut each branch separately to compromise with the limited size of the original A4 transparent sheets.
The typical pattern of a single branch is shown in fig.~\ref{fig:DiagramCoupled}~(a).
Once again, we used double-sided tape and pre-cut tabs to generate a single coherent structure.
To follow as closely as possible our previous analysis, we folded an origami bellows with five branches and four barrels, with pattern angles $\phi_1 = [81^\circ,79^\circ,77^\circ,75^\circ]$.
As for the stack of modules presented in the main paper, we tested the transition diagram for the system through quasi-static actuation.
The result is shown in fig.~\ref{fig:DiagramCoupled}~(b).

At first glance, the diagram seems identical to the Preisach graph presented in section~\ref{subsec:decoupled:bellows} and follows the same hierarchical structure of transitions.
Yet, some reachable states are defective configurations.
Furthermore, we observe transitions already defined in the context of GoE microstate transition with \emph{shifting} and \emph{erasure} of the defect.
But we also observe transitions that we name \emph{appearance} where one degree of defect is added to the system.
These three types of transitions round up the necessary framework to analyze the transformation of defective structures.

This specific diagram is actually easy to understand empirically.
We observe the appearance of a defect only when the third barrel, corresponding to $\phi_1 = 77^\circ$ , transition from {\tt 1} to {\tt 0}.
If the configuration allows it, {\it i.e.} if at least another barrel is deployed, the third barrel prefers to transition to the {\tt -} state and create a defect.
Finally, the system remains with a defect until either it has to be folded entirely ({\tt 0000}) or the three softer barrels have to be deployed ({\tt 0111} or {\tt 1111}).
Meanwhile, the defect shifts and the other bellows deploy and fold according to the mechanical rules established in section~\ref{subsec:mechanics_goe}.

The appearance of defects can be attributed to two elements.
First, it might simply come from an error during production.
Then, the analysis of the transition diagram is an interesting tool to understand where the irregularity lies.
For instance, a misplacement of the double-sided tape could lead to one-half of the third barrel having difficulty folding.
A second possibility comes from the difference in the cutting pattern.
Contrary to the modular approach, the barrels in a multi-barrel design are linked by a single crease with an angle that depends on the state of the two sections it connects.
The angle ranges from $\pi$, if both connecting sections are deployed, to 0, if they are folded.
An interesting case emerges when only one side is folded.
The crease's stiffness may change the stability of the structure, effectively changing the mechanical properties of each module depending on the states of its neighbors.
As a result, a destabilization of the non-defective relative to the defective configurations facilitates the formation of defects and explains why we manage to reach them quasi-statically.
We leave an investigation of the relation between properties of the connecting crease and change of the transition diagram to future studies.

\newpage
\end{document}

%% file: UnitCellGeometry.tex
\tikzmath{
\done=2.5;
\phione=75;
\phitwo=30;
\h=\done/sin(\phione-\phitwo);
\ltwo=\h*sin(\phitwo);
\lthree=\h*sin(\phione);
\dx=\ltwo*cos(\phione);
\xbound=0.5*\done;
\vecl=0.4;
\pentscale = 0.6;
\pentleg = \pentscale*\h;
\midpent = tan(54)*0.5*\pentleg;
}
\begin{tikzpicture}[every path/.style={line width = 1pt}]


\foreach \pm in {-1,1}{
\begin{scope}[yscale=\pm]
\path
coordinate (p1) at (0:0)
coordinate (p3) at ++(\phione:\ltwo)
coordinate (p2) at +(0:\done)
coordinate (p4) at (\phitwo:\lthree);

\draw[line join=bevel, color = NCSUk25] ([xshift = -\done cm]p1)--([xshift = -\done cm]p2)--([xshift = -\done cm]p4)--cycle;

\draw[line join=bevel, color = NCSUk25] (p1)--(p4)--(p3)--cycle;
\draw[line join=bevel, color = NCSUk25] (p1)--(p2)--(p4);
\end{scope}
}

\draw[-latex, color = BioIndigo, thick] (p1)--++(-\phione:\ltwo/2) node[at end, below=0.02cm ,inner sep = 0pt, fill=white] {$\bm{\omega}_2'$};
\draw[-latex, color = BioIndigo, thick] (p1)--++(-\phitwo:\lthree/2) node[at end , below,inner sep=0pt, fill=white] {$\bm{\omega}_3'$};

\draw[->, color = black] (p1)+(-\vecl,0) arc (180:\phione:\vecl) node[near end,left, inner sep = 0pt, fill = white] {\small $\pi\!-\!\phi_1$};

\draw[->, color = black] (p1)+(2*\vecl,0) arc (0:\phitwo:2*\vecl) node[near end,right=0.05cm, inner sep = 0pt, fill = white] {\small $\phi_2$};

\draw[-latex, color = black, thick] (p1)--++(0:\xbound) node[at end, right=0.02cm, inner sep = 0pt, fill=white] {$\bm{\omega}_1$};
\draw[-latex, color = black, thick] (p1)--++(\phione:\ltwo/2) node[at end, above=0.02cm ,inner sep = 0pt, fill=white] {$\bm{\omega}_2$};
\draw[-latex, color = black, thick] (p1)--++(\phitwo:\lthree/2) node[at end , above,inner sep=0pt, fill=white] {$\bm{\omega}_3$};
\draw[-latex, color = black, thick] (p1)--++(180:\xbound) node[at end,left, inner sep = 0pt, fill=white] {$\bm{\omega}_4$};

\pgfmathsetmacro{\ehh}{\lthree*sin(\phitwo)}
\draw[|<->|, color = black] (p4)++(0,0)--++(-90:\ehh) node [pos = 0.65, fill=white, inner sep = 1pt] {$h$};

\draw[color = ReynoldsRed,dashed] (-\xbound,0)--++(\phione:\ltwo/2)--++(\done,0)--++(\phione:-\ltwo/2);

\node[scale=1.5, inner sep = 2pt, anchor = north west] at (-\done,0.5*\h) {(a)};

\end{tikzpicture}\quad
\begin{tikzpicture}[every path/.style={line width = 1pt}]
\node[scale=1.5, inner sep = 2pt, anchor = north west] at (2.5*\xbound,0.5*\h) {(b)};

\coordinate (psio) at (3.3*\xbound,-.4*\h);
\draw[-] (psio)--++(0:0.7*\done) node[at end, anchor = west] {y};
\draw[-] (psio)--++(90:\done) node[at end, anchor = south] {z};
\filldraw (psio) circle (0.07);
\draw (psio) circle (0.15);
\node[anchor = east, inner sep = 0.2cm] at (psio) {x};

\draw[-latex] (psio)--++(40:\ltwo) node[at end, anchor = west, inner sep = 0.5pt] {$\bm{\omega}_3$};
\draw[dotted] (psio)+(40:\ltwo) arc (40:90:\ltwo) node[midway, fill = white, inner sep = 1.5pt] {$\Psi$};

\end{tikzpicture}\quad
\begin{tikzpicture}[every path/.style={line width = 1pt}]
\filldraw[color = white] (-0.35*\pentleg,-.1) rectangle ++(1.7*\pentleg,\h);

\node[scale=1.5, inner sep = 2pt, anchor = north west] at (-0.35*\pentleg,\h-.1) {(c)};

\draw[line width = 0.75pt, NCSUk60] (\pentleg,0)--++(72:\pentleg)--++(144:\pentleg)--++(216:\pentleg)--++(288:\pentleg)--cycle;
\draw[latex-latex] (0:0.5*\pentleg)-- (0,0) -- (108:0.5*\pentleg) node[at end, anchor = south west, inner sep = 1pt] {$\bm{\omega}_4$};
\node[anchor = south west, inner sep =1pt] at (0:0.5*\pentleg) {$\bm{\omega}_1$};

\draw[dotted] (0.3*\pentleg,0) arc (0:108:0.3*\pentleg) node[midway, fill = white, inner sep= 0.5pt, anchor = south west] {$\theta$};

\coordinate (ori) at (0.5*\pentleg, \midpent);
\draw[-] (ori)--++(0:0.3*\pentleg) node[at end, anchor = west] {x};
\draw[-] (ori)--++(90:0.3*\pentleg) node[at end, anchor = south] {y};
\filldraw (ori) circle (0.07);
\draw (ori) circle (0.15);
\node[anchor = east, inner sep = 0.2cm] at (ori) {z};

\end{tikzpicture}